%% file: ms_final.tex
\documentclass[%
preprintnumbers,
twocolumn,
 amsmath,amssymb,
 aps,
prc,
]{revtex4-1}
\usepackage[dvipsnames,usenames]{xcolor}
\usepackage{graphicx}
\usepackage{dcolumn}% Align table columns on decimal point
\usepackage{bm}% bold math
\usepackage{mathrsfs,natbib}
\usepackage{graphicx,epsf,amssymb,amsbsy,amsfonts,amssymb,amsmath}
\usepackage{slashed}
\usepackage{comment}
\usepackage{ulem}
\usepackage{hyperref}% add hypertext capabilities
\newcommand{\be}{\begin{equation}}
\newcommand{\ee}{\end{equation}}
\newcommand\beq{\begin{eqnarray}}
\newcommand\eeq{\end{eqnarray}}

\newcommand{\gcm}{{\rm g/cm}^3}
\newcommand{\ye}{{\rm Y_e}}

\newcommand{\nn}{\nonumber}

\begin{document}

\preprint{INT-PUB-17-045, LA-UR-17-29841}

\title{Nuclear pasta in hot dense matter and its implications for neutrino scattering}% Force line breaks with \\
\author{Alessandro Roggero}%
\email{roggero@lanl.gov}
\affiliation{
Institute for Nuclear Theory, University of Washington, Seattle, WA 98195  \\
Theoretical Division, Los Alamos National Laboratory, Los Alamos, New Mexico 87545, USA
}%[

\author{J\'er\^ome Margueron}
\email{jmargue@uw.edu}
\affiliation{%
Institute for Nuclear Theory, University of Washington, Seattle, WA 98195  \\
Institut de Physique Nucl\'eaire de Lyon, CNRS/IN2P3, Universit\'e de Lyon, Universit\'e Claude Bernard Lyon 1, F-69622 Villeurbanne Cedex, France
}%
\author{Luke F Roberts}
\email{robertsl@nscl.msu.edu}
\affiliation{
National Superconducting Laboratory \& Department of Physics and Astronomy, Michigan State University, East Lansing, MI 48824, USA
}%
\author{Sanjay Reddy}
\email{sareddy@uw.edu}
\affiliation{
Institute for Nuclear Theory, University of Washington, Seattle, WA 98195
}%

\date{\today}% It is always \today, today,

\begin{abstract}
We find that the abundance of large clusters of nucleons in neutron-rich matter
at sub-nuclear density is greatly reduced by finite temperature effects when
matter is close to beta-equilibrium. Large nuclei and exotic non-spherical
nuclear configurations called pasta, favored in the vicinity of the transition
to uniform matter at $T=0$, dissolve at relatively low temperature. For matter
close to beta-equilibrium we find that the pasta melting temperature
is $T_m^\beta\simeq  4\pm 1$~MeV for realistic equations of state.  The mechanism
for pasta dissolution is discussed, and in general $T_m^\beta$ is shown to be
sensitive to the proton fraction.  We find that coherent neutrino scattering from nuclei
and pasta makes a modest contribution to the opacity under the
conditions encountered in supernovae and neutron star mergers.  Implications for
neutrino signals from galactic supernovae are briefly discussed.
\end{abstract}

%\pacs{Valid PACS appear here}

\maketitle

\section{Introduction}
The properties of hot dense matter encountered in core-collapse supernovae,
newly born neutron stars called proto-neutron stars, and in neutron star mergers
is expected to play a key role in shaping their observable photon, neutrino and
gravitational wave emission. In supernovae,
state of the art simulations indicate that neutrino transport at high density
influences the supernova mechanism  \cite{Burrows:2012,JankaReview:2012}, the
long term neutrino emission detectable in terrestrial neutrino detectors
\cite{Burrows:1986me,Pons1999,Fischer2011,Roberts2016}, and heavy element
nucleosynthesis \cite{Hudepohl:10, Fischer:11, Roberts:12, Martinez-Pinedo:12,
Wanajo:13}. 

The presence of heterogeneous matter at high density is
expected to modify the neutrino scattering rates because the size of structures
encountered in such matter can be comparable to the neutrino wavelength, and
neutrinos would couple coherently to the net weak charge contained within them.
A familiar example is neutrino-nucleus coherent scattering, known to play an
important role in trapping neutrinos during core-collapse
\cite{Freedman:1973yd}. Additionally, heterogeneous phases are favored near
first-order phase transitions in neutron stars at high density \cite{Glendenning:1992vb}, and
coherent neutrino scattering in such matter can greatly increase the
opacity \cite{Reddy:1999ad}. Coherent neutrino scattering from the
nuclear pasta phase where large spherical and non-spherical nuclei coexist with
a dense nucleon gas for densities between $10^{13}-10^{14}$ g/cm$^3$ has also
been studied \cite{Horowitz:2004yf,Sonoda2007}. 

Recently, the enhanced neutrino opacity in the high density heterogenous pasta
phase was incorporated in simulations of proto-neutron star evolution and found
to have a significant impact on the temporal structure of the neutrino
luminosity \cite{Chuck2016}.  Motivated by this interesting finding, we perform
calculations of matter at finite temperature to  address if heterogeneous
nuclear pasta is present under the typical thermodynamic conditions  encountered
in proto-neutron stars, and study its influence on the neutrino scattering
rates.  We find that the heterogeneous pasta phase dissolves at relatively low
temperature for the small values of the electron fraction characteristic of
dense matter in beta-equilibrium.  Consequently, the enhancement of neutrino
scattering rates due to coherent scattering is relatively modest and
significantly smaller than those employed in \cite{Chuck2016}. In addition, we
find that Coulomb correlations between clusters  suppresses scattering of
neutrinos with wavelengths larger than the inter-cluster distance in agreement with 
earlier work \cite{Itoh1975,Itoh2004}. Interestingly, we also find that at lower temperatures when large nuclei can be 
present there could be a net reduction of the neutrino opacity as nucleons get locked up
inside nuclei. 

The material is organized as follows. In \S \ref{sec:matter} we discuss the
basic nuclear physics of phase coexistence and show that the simplified Gibbs
construction for two-phase equilibrium provides a useful bounds on the phase
boundaries between homogeneous and heterogeneous matter. This allows us to
provide an upper limit on the critical temperature above which pasta dissolves
to form a uniform nucleon liquid, and its dependence on the nuclear equation of
state is discussed. Implications for neutrino transport in proto-neutron stars
are discussed in \S \ref{sec:neutrino},  and our conclusions are presented in \S
\ref{sec:conclude}.

\section{Hot matter at sub-nuclear density and the dissolution of pasta}   
\label{sec:matter} 

The structure of matter at sub-nuclear density and at zero temperature is fairly
well understood. With increasing density, nuclei become neutron-rich due to the
rapid increase in the electron Fermi energy.  Neutrons drip out of nuclei when
the density exceeds $\rho_{\rm drip} \simeq 4  \times 10^{11}~ \gcm$, and
non-spherical or pasta nuclei are likely when the density  exceeds $\rho_{\rm
pasta} \simeq 10^{13}~ \gcm$  \cite{Pethick:1995di}. Several studies using
different many-body methods and underlying nuclear interactions have all yielded
similar qualitative behavior~\cite{Pethick:1995di,Douchin2001,Pearson2012}. 

There also
exist some calculations at finite temperature which indicate that at the highest
densities nuclei and pasta persist up to $T\simeq 10-15$ MeV when the electron
(or proton) fraction $\ye \gtrsim 0.1$ \cite{PhysRevC.79.055801,Avancini2008}.
In what follows we shall derive an upper bound on the temperature for the
dissolution of nuclei and pasta for beta-equilibrated matter at densities in the
range $\rho \simeq 10^{12} -10^{14} ~\gcm$. We consider beta-equilibrium matter
because the outer regions of a proto-neutron star, which may contain nuclear
pasta, are able to deleptonize rapidly and therefore reach beta-equilibrium on a
short time scale. 

First, we identify the thermodynamic conditions favorable for the existence of 
nuclear pasta. Since surface and Coulomb energies act to disfavor the heterogeneous state, and 
shell effects are relatively small at the temperatures of interest, the liquid-gas phase coexistence region
predicted by the Gibbs construction, where these effects are ignored,  will likely enclose 
the phase coexistence region predicted when such finite size effects are included. 
This simple observation allows us to provide 
a useful upper bound on the melting temperature 
by examining the two-phase Gibbs construction for bulk matter.  

For nuclei or pasta to coexist with a gas of nucleons, the high density liquid
phase inside these structures have to be in equilibrium with the low density gas
outside. Denoting the pressure, and the neutron and proton chemical potentials
of the high density liquid phase as $P^h$, and $\mu^h_n$ and $\mu^h_p$,
respectively, Gibbs equilibrium requires  $P^h=P^l$, $\mu^h_n=\mu^l_n$ and
$\mu^h_p=\mu^l_p$, where $P^l$, $\mu^l_n$ and $\mu^l_p$, are the corresponding
pressure and chemical potentials in the low density gas phase. To find the
coexistence region in the phase diagram an equation of state which specifies how
the energy density of bulk nucleonic matter $\varepsilon_{\rm nuc}(n_n,n_p,T)$
depends on the neutron and proton densities, and the temperature is needed. 
In practice we work in the proton-canonical ensemble where $\mu_n$ is fixed
and $n_p$ is the extensive variable~\cite{Ducoin2006}. 
We have however checked that our results are independent of the statistical ensemble.

\begin{figure}[tb] %  figure placement: here, top, bottom, or page
\centering
\includegraphics[width=0.99\linewidth]{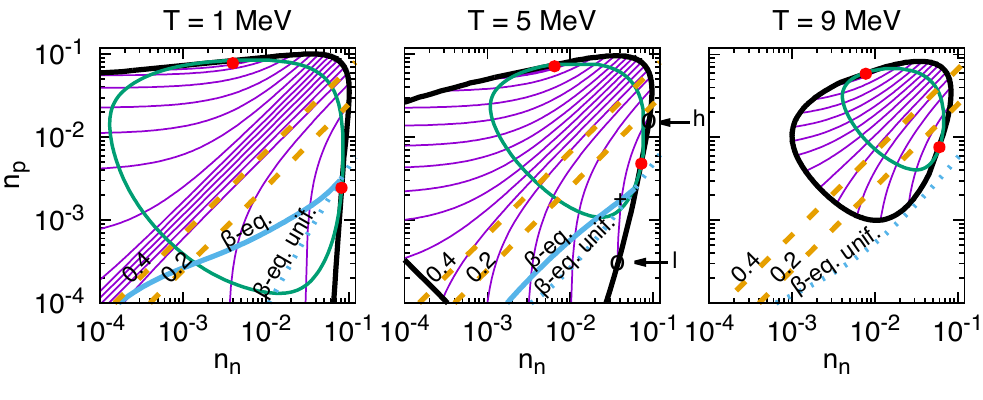} 
\caption{$\beta$-equilibrium path (blue line) for SLy4 Skyrme interaction in the coexistence region (delimited by the solid black symbols) in
the ($n_n$, $n_p$) plane and for $T$=1--9~MeV.
A sample of the Gibbs construction paths are shown (purple lines). 
The global density which intersect the $\beta$-equilibrium path in blue solid line with the Gibbs construction one (see for instance the "+"symbol at T=5 MeV), 
represents the equilibrium state connecting the 2 phases at equilibrium which are located at the boundaries (see the points labelled l and h which are the low and high density phases associated to the global density identified by the "+" symbol at $\beta$-equilibrium).
The constant $Y_e=0.2$, 0.3 and 0.4 paths are represented by the yellow dashed lines.
The spinodal instability region is also shown (green line) and the critical points are shown with red solid dots with error-bars.}
\label{fig:coexist}
\end{figure}

At a fixed temperature, phase coexistence is possible when there exists two
pairs of nucleon densities, denoted by $ n^h_n,n^h_p$ and $ n^l_n,n^l_p$, that
can satisfy the Gibbs equilibrium criteria. These pairs can be depicted as two
points on a two-dimensional plot where the axes are neutron and proton
densities. In Fig.~\ref{fig:coexist} these points are calculated for the model
SLy4 and appear on the solid-black curve. For a pair of points in Gibbs
equilibrium, a Gibbs construction can be used to find the state of matter at
intermediate densities. Therefore, a pair of points that satisfy Gibbs
equilibrium define a curve through neutron-proton density space given by  
\begin{eqnarray}
n_n = u n_n^h + (1-u) n_n^l \nonumber \\
n_p = u n_p^h + (1-u) n_p^l,
\end{eqnarray}
where $u$ is fraction of the volume that is occupied by the high-density liquid
phase. The purple curves represent these curves for pairs of select Gibbs equilibrium points.
For example, in the middle panel of Fig.~\ref{fig:coexist}, the pair of end
points defined by the intersection labeled $l$ and $h$ specify the neutron and
proton densities of the low and high density phases, $n^l_n,n^l_p$ and $
n^h_n,n^h_p$, respectively. Clearly, $Y_e$ varies along any Gibbs construction
curve, so a constant $Y_e$ curve crosses the Gibbs constructions of many Gibbs
equilibrium pairs in the mixed-phase region.  In Fig.~\ref{fig:coexist}, the
yellow dashed lines show curves of constant $Y_e$ and the Gibbs equilibrium at a
specific $Y_e$ is defined by its intersection with the magenta curve. The
solid-blue curve denotes the $\beta$-equilibrium path, along which
$\mu_n-\mu_p=\mu_e$. Gibbs equilibrium is possible along the $\beta$-equilibrium
path when solid-blue curve lies within the coexistence region. Once again, it
can be seen that the $\beta$-equilibrium curve moves across many Gibbs equilibrium
pairs as it traverses the coexistence region. The $\beta$-equilibrium path for
the homogeneous phase is also shown as the dashed-blue curve for reference. The
spinodal region where matter is unstable to small density perturbations is the
region enclosed by the green curve, and the critical points associated with the
first-order transition are denoted by  the red dots.  

Several insights about the role of finite temperature can be gleaned from
examining the progression of the phase coexistence region with temperature seen
in the three panels in Fig.~\ref{fig:coexist}:  
\begin{itemize} 
\item  With increasing temperature the extent of the phase co-existence region shrinks, and its intersection with the path of $\beta$-equilibrium decreases.  Above the critical temperature, $T^\beta_{\rm max}$ ($\simeq 9$MeV for the model chosen) there is no intersection and phase coexistence in $\beta$-equilibrium is not possible.

\item In contrast, out of $\beta$-equilibrium for moderate values of $Y_e > 0.2$
there exists a range of ambient conditions that extends to higher temperature
where Gibbs equilibrium is possible. Nonetheless, with increasing temperature
the area enclosed by the solid-black coexistence curve shrinks and its
intersection with lines of constant $Y_e$ is reduced.  Eventually, above the
critical temperature denoted by $T^{Y_e}_{\rm max}  \simeq 12-15$ MeV there is no intersection 
and phase coexistence is absent.   

\item Co-existence in $\beta$-equilibrium ends near the critical point. With
increasing temperature, phase co-existence ends by making a transition to the
uniform low-density gas phase. This feature, called retrograde condensation
\cite{Muller:1995ji}, implies that the path along
beta-equilibrium will favor fewer nuclei with increasing density.  

\item For moderate values of $Y_e > 0.2$ phase co-existence ends by transiting
to the high-density liquid phase and large nuclei persist to higher temperature. 

\item With increasing temperature, the density contrast between the high and low
density phases associated with Gibbs equilibrium is reduced. 
\end{itemize}

\begin{figure}[t] %  figure placement: here, top, bottom, or page
%\centering
\includegraphics[width=0.99\linewidth]{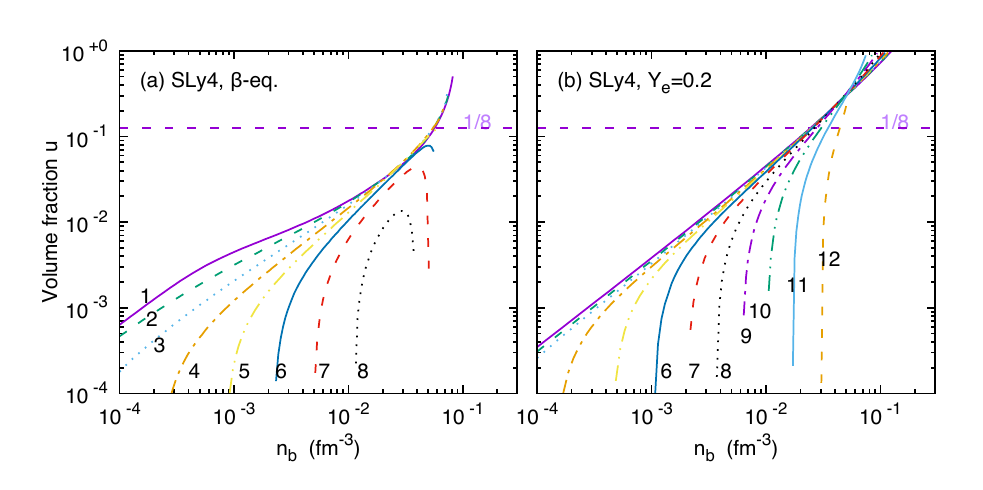} 
\caption{Volume fraction of the high density phase in heterogenous matter for SLy4 Skyrme interaction. (a) for the $\beta$-equilibrium path, (b) for the constant $Y_e=0.2$ path.}
\label{fig:vfrac}
\end{figure}

The impact of retrograde condensation on the volume fraction of the high-density
liquid phase is seen more clearly in Fig.~\ref{fig:vfrac}. At low
temperatures, $u$ begins close to zero at low densities and increases to one at high
densities, implying that it exits the coexistence region in the high-density
phase. But above a critical temperature, $u$ reaches a maximum of less than one
and turns over, implying the $\beta$-equilibrium path exits the coexistence
region in the low-density gas phase. The fact that the maximum volume fraction
occupied by the high-density phase, which corresponds to nuclei or pasta
structures, is rather small at temperatures high enough for retrograde
condensation can significantly impact the contribution of coherent scattering to
the neutrino opacity of $\beta$-equilibrium matter. Since non-spherical shapes
or pasta nuclei are favored for $ u \gtrsim 1/8$ (for a pedagogic discussion of
pasta nuclei see Ref.~\cite{Pethick:1995di}) we include the horizontal dashed line at
$u=1/8$ to help extract the critical temperature $T^\beta_m$ above which pasta
nuclei no longer appear (note that $T^\beta_m < T^\beta_{\rm max}$).  From panel
(a) of Fig.~\ref{fig:vfrac} we see that $T^\beta_m$ is between 5 and 6~MeV (for
SLy4 EOS). In contrast for matter at fixed $Y_e=0.2$, shown in panel
(b), pasta nuclei persist to higher temperatures until phase coexistence
ends at  $T^{Y_e}_{\rm max}$

\begin{figure}[tbp] %  figure placement: here, top, bottom, or page
\centering
\includegraphics[width=0.99\linewidth]{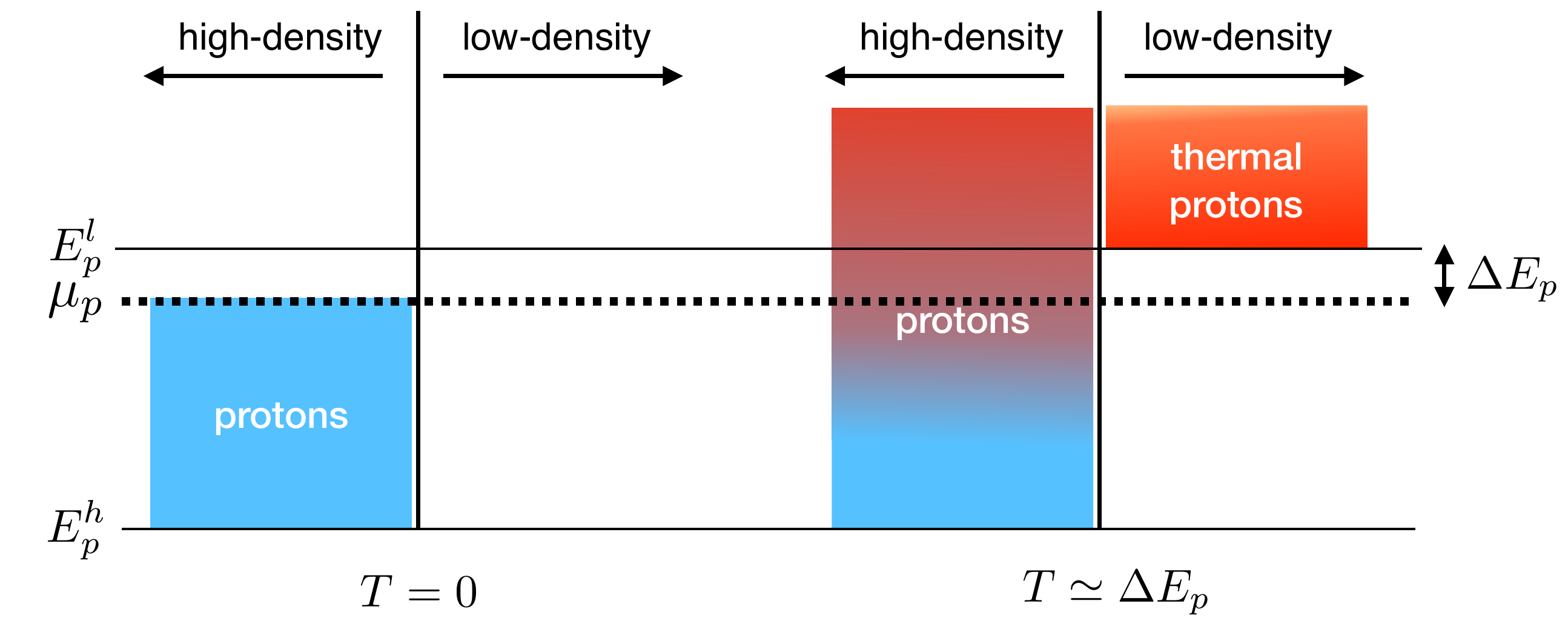} 
\caption{This schematic shows the occupied proton energy levels in the low and high density phases that coexist in the heterogeneous phase. The $T=0$ situation is shown on the left and finite temperature where a significant thermal population exists in the low density phase is on shown the right. See text for additional details.}
\label{fig:thermal_protons}
\end{figure}

We can understand the physical mechanism for retrograde condensation at larger
temperatures by
examining the evolution of the proton fraction in the gas phase.  Global charge neutrality
requires the volume fraction of the high density phase to be
\be
u  = \frac{n_e-n^l_p}{n^h_p-n^l_p }\,,
\label{eq:vfrac}
\ee    
where the electron density $n_e$ is assumed to be uniform because the Debye screening length is large compared to the typical size of electrically neutral Wigner-Seitz cells. In the beta-equilibrium mixed phase the lowest energy level for protons in the low density gas phase $E^l_p > \mu_p$ and at $T=0$ the proton density there denoted by  $n^l_p=0$.  At $T=0$ the volume fraction $u = n_e/n^h_p$ increases rapidly with increasing density because  $n_e$ increases and $n^h_p$ decreases. At finite temperature $n^l_p > 0$ because proton states in the gas can be thermally populated. This is illustrated in Fig.~\ref{fig:thermal_protons} where the occupied energy levels of protons in both the low and high density phases are shown at zero and finite temperature. 

The thermal population of protons in the gas 
\be 
n^l_p\simeq 2 \left( \frac{m_p T}{2\pi} \right)^{3/2} e^{-\Delta E_p/T}
\label{eq:nlp}
\ee 
where $\Delta E_p=E^l_P-\mu_p$, becomes significant when $T\simeq \Delta E_p$ and
increases exponentially with temperature. In contrast, the density of protons in
the high density phase still remains significantly larger and does not change
appreciably with temperature because of their high degeneracy.

\begin{figure}[tbp] %  figure placement: here, top, bottom, or page
\centering
\includegraphics[width=0.99\linewidth]{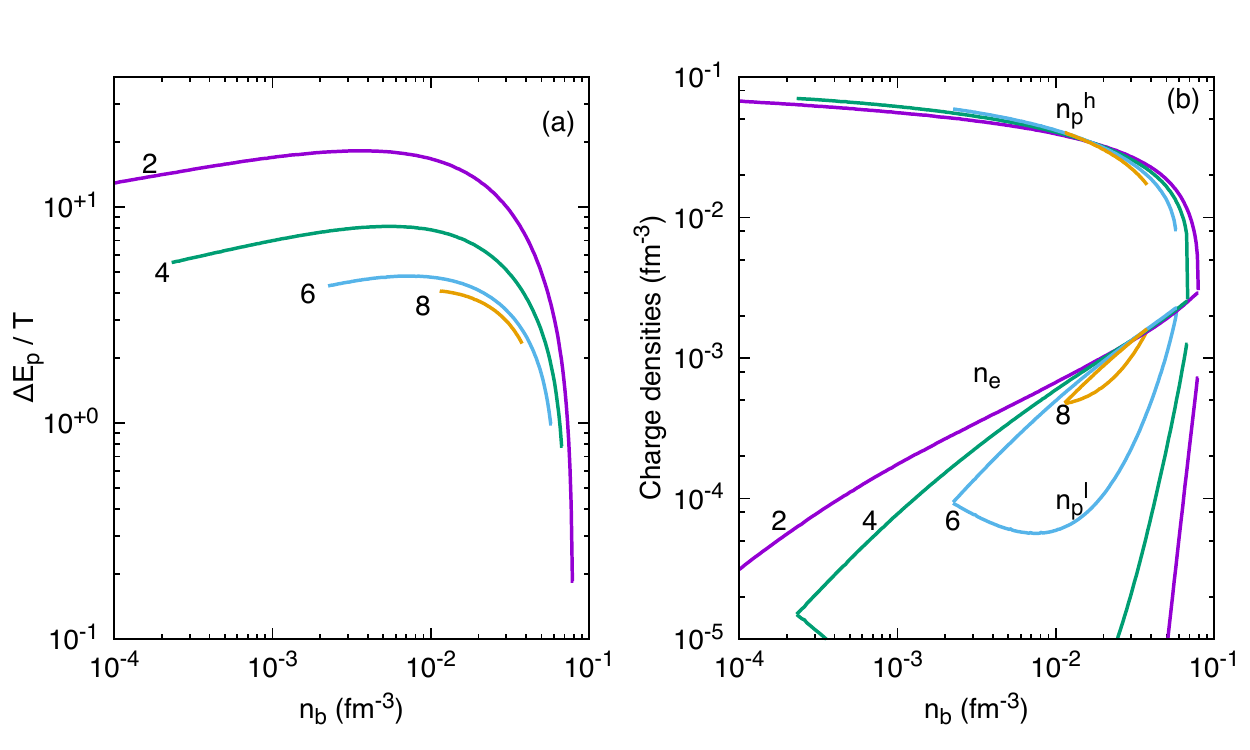} 
\caption{(a) Energy differences $\Delta E_p/T$ and (b) charge particle densities ($n_e$, $n_p^l$, $n_p^h$) as function of the density and for different temperatures $T=1$-6~MeV. 
The SLy4 EOS is considered here.}
\label{fig:de}
\end{figure}

The typical evolution of $\Delta E_p/T$ is shown in panel (a) of
Fig.~\ref{fig:de} for the SLy4 EOS. Except close to the transition density,
$\Delta E_p/T \gg 1 $ leads to significant suppression of the proton density in
the gas phase. In the vicinity of the transition density $\Delta E_p/T $
decreases rapidly and from Eq.~(\ref{eq:nlp}) the proton fraction in the gas
increases exponentially. The number densities of the charged particles as
function of the average baryon density are shown in panel (b). Since electric
charge neutrality in the uniform phase requires $n_e=n_p=n_p^{gas}$, the point
at  which $n_e$ and $n_p^{l}$ first intersect defines the low density boundary
of the coexistence region. In the vicinity of this point, nonuniform matter is
predominantly composed of the gas phase. The high density boundary is defined by
the intersection of the $n_e$ and $n_p^h$ at low temperature, or by the second
intersection of the $n_e$ and $n_p^{l}$ at high temperature as expected for
retrograde condensation.  These features are also readily discernible from
Fig.~\ref{fig:vfrac} where the evolution of the volume fraction of the high
density phase with density is shown for various temperatures. % Results for matter
%in $\beta$-equilibrium is shown in panel (a), and results for $Y_e=0.2$ and
%$Y_e= 0.4$ are shown in panels (b) and (c), respectively.  

As expected from the preceding discussion and Eq.~(\ref{eq:vfrac}), for matter
in $\beta$-equilibrium where $Y_e$ is small, the volume fraction $u$ will
decrease with density for $T\gtrsim \Delta E_p$. When this criterion is met, the density of protons in the low density gas phase will become comparable to the electron density, and eventually as  $\Delta E_p$ decreases with density the volume fraction $u \rightarrow 0$. 

We now turn to study the model dependence of the critical temperatures  denoted by $T^\beta_m$, $T^{\beta}_{\rm max}$,  and $T^{Y_e}_{\rm max}$ discussed earlier.   
\begin{figure}[t] %  figure placement: here, top, bottom, or page
%\centering
\includegraphics[width=0.99\linewidth]{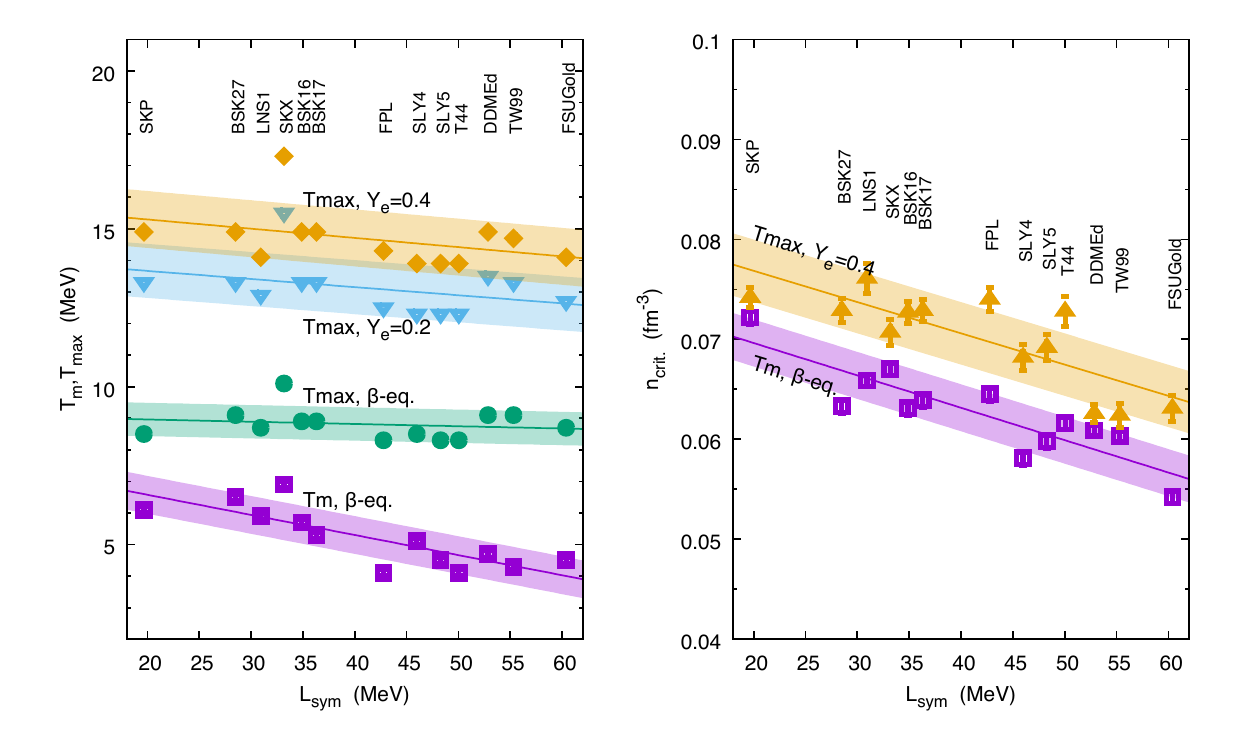} 
\caption{(a) Melting temperature $T_m$ and maximal temperatures $T_{max}$ at $\beta$-equilibrium, compared to $T_{max}$ at fixed $x_e$=0.2 and 0.4 for the set of EOS
compatible with chiral EFT EOS in neutron matter.
(b) Highest density reached by the pasta phase for beta equilibrium or at fixed $Y_e$.
These predictions are shown versus $L_{sym}$.}
\label{fig:TmvL}
\end{figure}
We first select a subset of model Skyrme and relativistic mean field EOSs which predict the energy per particle of neutron matter at $n_b=0.06$ and 0.10~fm$^{-3}$) that are compatible with QMC~\cite{Tews2016} or MBPT~\cite{Drischler2016}, which are based on two and three body chiral EFT potentials. The pasta melting temperature $T^\beta_m$ for these models are shown in panel (a) of Fig.~\ref{fig:TmvL}.  The names of the EOS are shown vertically above the predictions and the EOS are ordered according to the slope of the symmetry energy at nuclear saturation density denoted by $L_{sym}$. The average prediction is $T^\beta_m=5.0\pm2$~MeV and decreases with $L_{sym}$ (the anti-correlation coefficient is -0.81) and the dispersion reflects the additional dependence on the EOS parameters.  In panel (b) of Fig.~\ref{fig:TmvL} we show the highest average density of the coexistence region associated with $T^\beta_{m}$ and $T^{Y_e=0.4}_{max}$ for the EOSs in panel (a) and find that they are clearly anti-correlated with $L_{sym}$.

The striking feature here is that the pasta melting temperature in $\beta$-equilibrium is much lower than the maximal temperature of the phase coexistence, and that the maximal temperature $T^{Y_e}_{\rm max}$ even at a modest value of $Y_e=0.2$ is about a factor of two higher than $T^\beta_m$.  For typical values of $L_{sym}$ around 50--60~MeV, the melting temperature is estimated to be $T^\beta_m\simeq 4\pm1$~MeV. Since our analysis neglects finite size effects such as surface, Coulomb and shell effects we believe that this is an upper limit on the melting temperature.

\section{Neutrino scattering}
\label{sec:neutrino}

Coherent neutrino scattering from nuclei and pasta can be estimated using the
two-phase Gibbs construction discussed in the preceding section if their typical
size is known. The nuclear size is set by the competition between the surface
and Coulomb energies, the mass number and charge of the energetically favored
nuclei can be calculated by specifying the surface
tension~\cite{Pethick:1995di}. Shell effects can also play a role but we can
expect their impact to be less important at the temperatures of interest, and we
neglect them in the following analysis.  Further, although we should expect a distribution of nuclei at finite temperature, to obtain a simple first estimate we shall assume that the distribution is dominated by a single nucleus. In this case, the radius of the favored nucleus is  
\be 
r_{A}^3= \frac{3\sigma}{4\pi e^2 (n^h_p)^2  f_3(u)}  \; \hbox{ with } \;  f_3(u)= \frac{2-3u^{1/3}+u}{5} \,,  
\ee
where $\sigma$ is the surface tension between the low and high density phases,
and $f_3(u)$ is the geometrical factor associated with the Coulomb energy of the
Wigner-Seitz cell in $d=3$ dimensions \cite{Ravenhall:1983uh}. The surface
tension is a function of the density, $Y_e$ and $T$. We use the ansatz from
Ref.~\cite{Luke_surfacet} (see also Ref.~\cite{Lattimer1985}) and parameters
obtained for the SLy4 interaction. Note that this simple ansatz neglects the influence of the protons in the low density gas phase on the surface tension~\cite{Aymard2014} . 
 
For the purpose of calculating coherent neutrino scattering, we shall, for simplicity, assume that nuclei are spherical for all values of $u$. This is reasonable because angle averaged coherent scattering rates from rod-like and slab-like  structures have been calculated earlier and found to be comparable or smaller than those from spherical nuclei of similar size \cite{Alcain2017}.  Further, as noted earlier, close to $\beta$-equilibrium the pasta region is relatively small even for $T<T_m^\beta$, and absent for $T>T_m^\beta$.

The differential coherent elastic scattering rate from the nuclei in the heterogeneous phase is given by
\be 
\frac{d\Gamma_{\rm coh}}{d\cos{\theta}} = \frac{G^2_F~E^2_\nu}{8\pi}~n_A~(1+\cos{\theta}) ~S(q)~N^2_{\rm w}~F^2_A(q)
\ee
where the total weak charge of a nucleus is defined as
\be 
N_{\rm w} = \frac{4 \pi}{3}r^3_A~(n^h_n - n^l_n) \,,
\ee
and $n_A=3 u /(4\pi r^3_A)$ is the density of nuclei. We have neglected the proton contribution in the vector response because of their weak charge $\simeq 1-4 \sin^2{\theta_W} \approx 0$, and subtracted the density of neutrons from the low density phase because neutrinos only scatter off the density contrast. The static structure factor $S(q)$ accounts for correlations between nuclei due to long-range Coulomb interactions (weakly screened by electrons) that tends to suppress scattering at small momentum transfer $q = E_\nu\sqrt{2(1-\cos{\theta})} \lesssim1/a$ where $a=(3/4\pi n_A)^{1/3}=r_A/u^{1/3}$ is the average distance between nuclei. Scattering with high momentum transfer with $q \gtrsim 1/r_A$ is suppressed by the form factor of the nucleus $F_A(q)$ which we take to be that of a sphere of constant density and radius $r_A$. More realistic choices such as the Helm form factor~\cite{Helm}, have a negligible impact on our results. 

In a one component plasma, $S(q)$ depends on $a$ and the Coulomb coupling parameter $\Gamma =  Z^2e^2/a k_BT$ where $Z$ is the ion charge, $e^2=1/137$ and $k_BT$ is the thermal energy.  In our simple model for the heterogenous state where we assume a single spherical nucleus captures the essential physics 
\be
Z \approx  26 ~\left(\frac{\sigma}{\sigma_0}\right)~\left(\frac{n_0}{2 n^h_p}\right)~\left(\frac{f_3(0)}{f_3(u)}\right) \,,
\label{eq:simpleZ}
\ee
where $\sigma_0 \simeq 1.2 $ MeV/fm$^2$ is the surface tension of symmetric nuclei in vaccum, $n_0\simeq 0.16$ fm$^{-3}$ is the nuclear saturation density. Typically we find $Z \simeq 50$ at the density for which we expect an appreciable fraction of large nuclei or pasta, and $\Gamma \gg 1$. For large $\Gamma$ the static structure factor $S(q) \ll 1 $ unless $q a \gg 1$, and for $\Gamma > 10$ the interference of amplitudes for neutrino scattering off different clusters is strong and destructive at small $qa < 2\textendash3$. At intermediate $q a \simeq 4\textendash5$,  constructive interference can enhance scattering, and for $q a \gg 5$ where interference is negligible $S(q) \simeq 1$. In this work we employ $S(q)$ obtained from recent fits to accurate MD simulations of one-component plasmas~\cite{Desbiens2016} to properly account for screening for $\Gamma$ in the range $1\textendash150$. We note that for $T > 2$ MeV, $\Gamma  <150$ even at the highest density, and crystallization is not favored and its reasonable to work with $S(q)$ obtained for the liquid state. 

The neutrino scattering rate from non-relativistic nucleons in the gas phase is given by 
\beq 
\frac{d\Gamma_{\rm \nu}}{d\cos{\theta}} &=& \frac{G^2_F~E^2_\nu}{8\pi}~ \sum_{ij} \Big[ (1+\cos{\theta})~C^{i}_{v}C^{j}_{v} S^{ij}_v(q) \nn\\
&&\hspace{1.6cm}+  (3-\cos{\theta}) C^{i}_{a}C^{j}_{a} S^{ij}_a(q) \Big]\,,
\eeq
where the labels $i$ and $j$ can be either neutrons or protons, $C^i_v$ and $C^i_a$ are their corresponding vector and axial vector charges. In the long-wavelength limit, which is adequate to describe low energy neutrino scattering, the static structure factors (unnormalized) can be related to thermodynamic functions, 
\be 
S^{ij}_v =T \left(\frac{\partial^2 P}{\partial \mu_j \partial \mu_i}\right)_T 
\ee
where $P$ is the pressure of the gas phase and $\mu_i$ is the chemical potential of either neutrons or protons, and the axial or spin response  
\be 
S^{ij}_a =T \left(\frac{\partial^2 P}{\partial \delta_j \partial \delta_i}\right)_T\,, 
\ee
where $\delta_i$ is the chemical potential associated with the spin density of species $i$. When interactions between nucleons can be neglected, the structure functions 
greatly simplify and are given by  
\be 
S^{ij}_v =S^{ij}_a = \delta^{ij}  S_{gas}(\mu_i,T)\,,  
\label{eq:Sgas}
\ee
where 
\be 
S_{gas}(\mu_i,T)= \int \frac{d^3p}{(2\pi)^3}~\frac{e^{\beta(p^2/2m-\mu_i)}}{(1+e^{\beta(p^2/2m-\mu_i)})^2} \,,
\ee 
where $\beta=1/T$ and only correlations due to Fermi statistics are included. Strong nuclear interactions induce additional correlations between nucleons in the gas and can alter the structure factors. At the sub-nuclear densities of interest, calculations suggest a modest enhancement of the vector response, and a suppression by up to 50\% of the axial response \cite{Iwamoto:1982zp,Burrows:1998cg,Reddy:1998hb,Horowitz:2006pj,Horowitz:2016gul}.  Since our primary interest here is to asses the role of coherent scattering, in what follows we shall neglect corrections due to strong interactions and use Eq.~(\ref{eq:Sgas}) to calculate the scattering rates in the gas phase. 

To asses the importance of coherent scattering from heavy nuclei in the heterogeneous phase we define the ratio
\begin{widetext}
\be 
\begin{split}
\label{eq:ratio}
{\cal R}&=\frac{\sigma_{het}^{\rm tran}(E_\nu)}{\sigma_{hom}^{\rm tran}(E_\nu)}=\frac{\sigma_{coh}^{\rm tran}(E_\nu)+\sigma_{gas}^{\rm tran}(E_\nu)}{\sigma_{hom}^{\rm tran}(E_\nu)}
=\frac{n_A N^2_{\rm w}\langle S_{cl} (E_\nu)\rangle + (1+5 (C^n_A)^2) S_{gas}(\mu^{het}_n,T) + 5 (C^p_A)^2 S_{gas}(\mu^{het}_p,T)}{ (1+5(C^n_A)^2) S_{gas}(\mu^{hom}_n,T) + 5 (C^p_A)^2 S_{gas}(\mu^{hom}_p,T) }
\,,
\end{split}
\ee
\end{widetext}
where $\sigma^{\rm tran}= \int d\cos{\theta}~(1-\cos{\theta}) ~d\Gamma/d\cos{\theta}$ is the elastic transport cross section per unit volume for neutrinos.  ${\cal R}$ is analogous to the parameter $\xi$  introduced in \cite{Chuck2016}, and quantifies the change in neutrino scattering rates in heterogeneous phase, where both coherent scattering from nuclei ($\sigma_{coh}^{\rm tran}$) and scattering from free nucleons in the gas phase contribute. The term $\langle S_{cl} (E_\nu)\rangle$ in the cross section from clusters indicates angle averaging of the corrections due to correlations and nuclear form factors,
\be
\langle S_{cl} (E_\nu)\rangle=\frac{3}{4} \int_{-1}^{1} d~\cos{\theta} (1-\cos{\theta})(1+\cos{\theta})S(q)~F_A^2(q)
\ee
and is a function of $E_\nu$ trough $q=E_\nu\sqrt{2(1-\cos{\theta})}$. We note that neglecting both the correlations in the gas and the protons has a small impact of $\lesssim 10\%$ on the ratio. However, a strong suppression of the nucleon axial response due to spin correlations would reduce the opacity of the homogeneous phase, and favor larger  ${\cal R}$. 

\begin{figure}[t] %  figure placement: here, top, bottom, or page
%\centering
\includegraphics[width=0.99\linewidth]{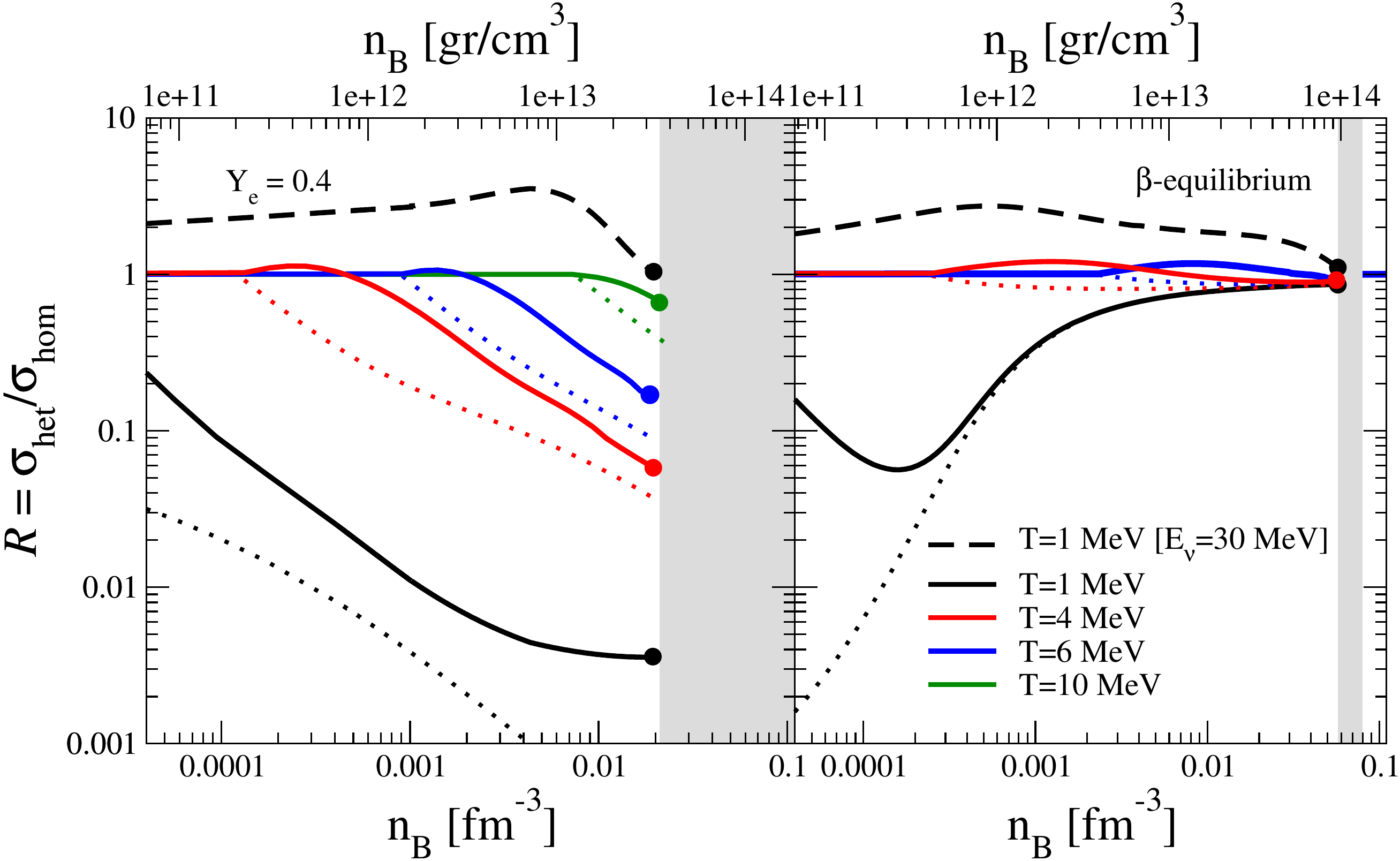} 
\caption{Ratio ${\cal R}$ from Eq.\eqref{eq:ratio} in two different conditions: fixed proton fraction $Y_e=0.4$ (left panel) and $\beta$-equilibrium (right panel). The gray band indicates the region in density where non-spherical pasta may be present, and the curves terminate with a dot for the density at which $u=1/8$. Dotted curves indicate the contribution of the external gas only. Neutrino are considered at thermal equilibrium, except for the black dashed curves where $E_\nu=30$~MeV.}
\label{fig:ratio}
\end{figure}

\begin{figure}[t] %  figure placement: here, top, bottom, or page
%\centering
\includegraphics[width=0.99\linewidth]{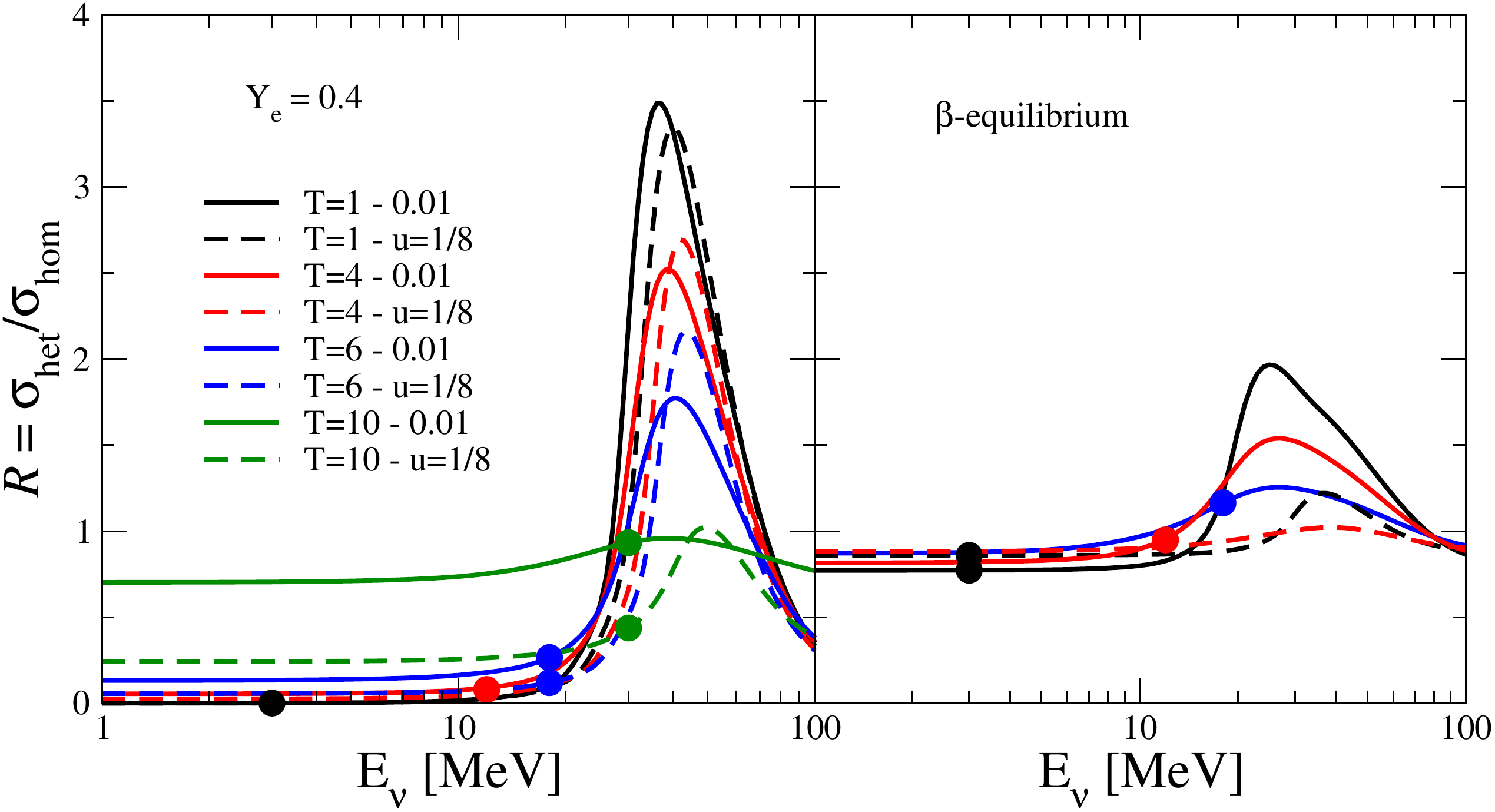} 
\caption{Energy depence of the total ratio of cross section ${\cal R}$ at various temperatures and for two different densities: $n_B=0.01$ fm$^{-3}$ (full lines) and the threshold density for which $u=1/8$ (dashed lines). As for Fig.\ref{fig:ratio} the left panel is for fixed proton fraction $Y_e=0.4$ and the right panel is for $\beta$-equilibrium conditions.
In both cases the full dots indicate the energy for thermal neutrinos ($E_\nu=3T$). }
\label{fig:ratio_ofE}
\end{figure}

The results for the ratio of cross section ${\cal R}$ are displayed in
Fig.~\ref{fig:ratio}. The left panels show results at fixed proton fraction
$Y_e=0.4$,  on the right panel results for matter in $\beta$-equilibrium are
shown. In both cases, with the exception of the black dashed line, neutrinos are
assumed to be thermal and their energy $E_\nu = 3 T$. The energy dependence of
the cross sections is shown in Fig.~\ref{fig:ratio_ofE}. The strong suppression
of coherent scattering at low energy is clearly visible, and the dot on each
curve corresponding to $E_\nu=3T$ shows that Coulomb correlation suppresses
scattering for neutrino energies of interest. The Coulomb parameter $\Gamma$ for
the plots in Fig.~\ref{fig:ratio_ofE} range from $\Gamma_{\rm min}=4 (6)$ for
$n_B=0.01$ fm$^{-3}$ and $T= 10 (6)$ MeV to $\Gamma_{\rm max}=150 (74)$ for
$u=1/8$ and $T=1$ MeV at fixed proton fraction (beta-equilibium). The value of
$\Gamma$ at select points is shown in Table I. At the lowest temperature of $T=1$ MeV and large proton fraction $Y_e =0.4$, our simple ansatz in Eq.~\ref{eq:simpleZ} predicts a large $Z > 60$ and $\Gamma > 200$. At these very low temperatures, it would be appropriate to use $S(q)$ from simulations of the solid phase. 
However, here we adopt the approximate treatment suggested in earlier studies
\cite{Chuck1997,Itoh2004} where they circumvent the problem by limiting the
value of the Coulomb coupling to $\Gamma_{max}=150$, and is indicated by an
asterisk in Table I.  These low temperature conditions are encountered only at late times in the proto-neutron star phase when the neutrino luminosity is greatly reduced and undetectable even for close by supernovae in detectors such SuperKamiokande where energy threshold is about 5 MeV. Additionally, shell effects can be important in the determination of $Z$ at low temperature and smaller values of  $Z\approx40-50$ are obtained at $T=0$ \cite{Douchin2001,Onsi2008}. Nonetheless, we included these low temperature results, which despite the approximations mentioned, provide useful insights about trends and allow for comparison with earlier work.

\begin{table}[tb]
\setlength{\tabcolsep}{.1in}
\renewcommand{\arraystretch}{0.9}
\centering
\begin{ruledtabular}
\caption{Values of Coulomb coupling $\Gamma$ for Fig.~\ref{fig:ratio_ofE}. }
\begin{tabular}{lllll}
T & $\rho=0.01$ & $u=1/8$ & $\rho=0.01$ & $u=1/8$ \\
in MeV & $Y_e=0.4$ & $Y_e=0.4$ & $\beta$ & $\beta$ \\ \hline
$1$ &             150$^*$          &            150$^*$           &     72.7                  & 74.0 \\
$4$ &             69.8          &           138.8            &         14.2              & 4.0 \\
$6$ &             35.1          &           75.4            &          6.3             & -- \\
$10$ &             4.0          &            17.4           &         --              & --
\end{tabular}
\end{ruledtabular}
\end{table}

From Figs.~\ref{fig:ratio} and \ref{fig:ratio_ofE} we can draw the following conclusions: 
\begin{itemize}  
\item At low temperature when large nuclei are present and persist up to high density, the opacity to high energy neutrinos with  $E_\nu \gtrsim 4/a$ where $a$ the distance between nuclei is enhanced, but coherent scattering is greatly reduced for low energy thermal neutrinos due to Coulomb correlations between nuclei. We find a net reduction in the scattering rates in the heterogeneous phase because a large fraction of free nucleons are tied up inside nuclei. In the homogeneous phase these nucleons make a significant contribution to neutrino scattering because they couple to the axial current.    
\item In $\beta$-equilibrium coherent scattering makes a relatively small contribution to the total neutrino opacity for all temperatures of interest.  At low temperature, when nuclei and pasta are present, Coulomb correlations reduce coherent scattering, and at high temperature, pasta and large nuclei melt. We find that scattering off nucleons in the gas phase dominates unless nuclear correlations can greatly suppress the spin response of dilute nuclear matter.   
\item Large opacity due to coherent scattering reported in Ref.~\cite{Chuck2016} arose because the neutrino energy was chosen to be large to ensure that the suppression due to Coulomb correlations was  mild, and it was assumed that pasta nuclei would survive up to  $T \simeq 10$ MeV in matter close to $\beta$-equilibrium.     
\item Fig.~\ref{fig:ratio_ofE} illustrates that the heterogenous phases can act as a low-pass filter for neutrinos. In the diffusive regime the strong energy dependence of the neutrino cross-sections would imply non-linear thermal evolution where cooling would accelerate rapidly with decreasing temperature.     
\end{itemize}

These results have significant implications for the impact of coherent pasta
scattering on proto-neutron star cooling. In Ref.~\cite{Chuck2016}, it was shown that
if coherent scattering from nuclear pasta increases the neutrino opacity
relative to that of a homogeneous gas, pasta formation in the outer layers of
the proto-neutron star can trap neutrino energy for the first few seconds after
a succesful core collapse supernova explosion. This heat trapping causes the
temperature of the outer layers of the proto-neutron to increase until they
reach the pasta melting temperature. This heats up the entire region over which
neutrinos decouple from matter, increasing the average energy of neutrinos
escaping from the proto-neutron star. Additionally, the energy that is trapped
initially gets out at later times. Both of these effects contributed to a more
detectable late-time neutrino signal. A pasta melting
temperature of $10 \, \textrm{MeV}$ was used in their parameterized simulations, but
it was suggested that reducing the melting temperature of the pasta could reduce the
impact of the pasta on the neutrino signal. 

Here, we have found that the pasta melting temperature for $\beta$-equilibrated
matter is $T^\beta_m \approx 4 \pm 1 \, \textrm{MeV}$ and that the presence of a
high-density phase can reduce the neutrino opacity.  First, this implies that
even if coherent scattering from nuclear pasta increased the neutrino opacity,
the impact of pasta on the proto-neutron star neutrino signal would be smaller
than the impact predicted by Ref.~\cite{Chuck2016}, since nuclear pasta would be
present for a shorter portion of proto-neutron star cooling. The reduced
critical temperature would also cause a smaller perturbation in the temperature
gradient near the neutrino sphere, which would reduce the enhancement of the
neutrino luminosity even when the pasta is present. Second, we predict that
correlations among high-density structures act to reduce the neutrino opacity
for neutrinos with energies $\lesssim 4/a$, which is an energy scale that
	is often significantly above the thermal energy.  Therefore, the
	presence of pasta may allow the majority of thermal neutrinos to escape
	more easily and potentially speed up neutrino cooling, thereby reducing
	the late-time neutrino detection rate from a nearby supernova.
	\section{Conclusions} \label{sec:conclude} 

In this paper, we have analyzed the properties of the hot nuclear pasta phase and we have shown the large qualitative differences between matter in $\beta$-equilibrium and at modest electron fraction $Y_e > 0.2$. In beta-equilibrium, we find that pasta melts or dissolves at relatively low temperature, reducing drastically the volume fraction occupied by the large nuclei. With increasing temperature protons leak out of nuclei, enter the gas phase and alter the nature of the transition to bulk matter. Here, nuclei dissolve with increasing density in a phenomena referred to as retrograde condensation. We have introduced a new temperature, the pasta melting temperature $T_m$, above which the volume fraction of nuclei cannot exceed $1/8$. In $\beta$-equilibrium the melting temperature $T_m^\beta \simeq 4\pm1$~MeV for EOS with $L_{sym}=50-60$~MeV and compatible with EFT predictions in neutron matter. The melting temperature $T_m^\beta$ was found to decrease with increasing $L_{sym}$. For matter with $Y_e > 0.2$ large nuclei and pasta persist to higher temperatures $ T_m \simeq 15$~MeV and retrograde condensation is absent.  

In the second part of our paper, we have analyzed the impact of the coherent
scattering off nuclear clusters on the neutrino opacities, for thermodynamical
conditions corresponding to core-collapse supernovae or neutron star mergers. We
found that both the retrograde condensation and the Coulomb correlations in the
static structure factor contribute to reduce the impact of coherent scattering
on neutrino opacities.  For matter far out of beta-equilibrium where heavy
nuclei and pasta persist to high temperatures, Coulomb correlations between
clusters greatly reduce the coherent scattering rates at high density. Here,
rather than an increase, we found a net reduction in the opacity for thermal neutrinos when clusters are present. This may be important at very early times post bounce during the supernova when matter with large $Y_e$ is encountered briefly during the period when lepton number is trapped.  On longer timescales characteristic of proto-neutron star evolution, beta-equilibrium favors much smaller values of $Y_e$,  and for $T < T_m^\beta$ only a moderate increase by less than 20\% is found for thermal neutrinos, at variance with the factor 5 reported in Ref.~\cite{Chuck2016}. 
We find such an increase only for high energy non-thermal neutrinos, for which correlations between nuclei enhance the scattering rates. 

While we believe the physical effects mentioned above are robust, additional work is warranted to 
obtain more quantitative predictions. Hartree-Fock calculations, such as those being reported in 
Ref.~\cite{Newton:2009zz,Schuetrumpf:2016kzq} which self-consistently include the
surface tension, Coulomb, and shell effects, would provide improved estimates for $T_m^\beta$ to better constrain the
temperature range in which pasta is present. It will also be desirable to go
beyond the single-nucleus approximation in calculating the ion structure factor,
and include in addition non-spherical shapes. Ultimately, these modifications to
the neutrino opacities need to be incorporated self-consistently with the
underlying equation of state in proto-neutron star and supernova simulations to
asses if the presence of nuclear clusters at sub-nuclear density can influence
supernova observables. Nonetheless, it seems likely that retrograde condensation
and ion-correlations will together disfavor the large changes to the temporal
structure of the neutrino signal predicted in Ref.~\cite{Chuck2016}.

\section{Acknowledgements}
We would like to thank C. Horowitz and W. Newton for discussions. SR was supported by DOE Grant No. DE-FG02-00ER41132. AR was supported by NSF Grant No. AST-1333607 and by DOE Grant No. DE-AC52-06NA25396. LR was supported by the DOE Office of Science under Award No. DE-SC0017955. 
JM was partially supported by the IN2P3 Master Project MAC.
\bibliographystyle{apsrev4-1}
%\bibliography{ms-bib}
\input{ms_final-bib}

\end{document}